\documentclass{jkas}

\usepackage{xcolor}
\usepackage{multirow}
\usepackage{graphics}
\usepackage{longtable}
\usepackage{amsmath}
\usepackage{mathtools}
\usepackage{flushend}

\newcommand\tnote[1]{$^{\rm #1}$}

\def\year{2022} 
\def\month{December} 
\def\volume{55} 
\def\issue{6} 
\def\beginpage{1} 
\def\endpage{99} 
\def\received{October 13, 2022} 
\def\accepted{December 15, 2022} 
\date{Received \received; accepted \accepted}

\usepackage[colorlinks,
            pdftex,
            breaklinks,
            linkcolor=blue,
            urlcolor=blue,
            anchorcolor=blue,
            menucolor=blue,
            citecolor=blue]{hyperref}

\title{
Using the Crab Nebula as Polarization Angle Calibrator for the Korean VLBI Network 
}

\author[1]{\href{https://orcid.org/0000-0001-9799-765X}{Minchul Kam}}
\author[1,2]{\href{https://orcid.org/0000-0003-0465-1559}{Sascha Trippe}}
\author[3]{\href{https://orcid.org/0000-0003-1157-4109}{Do-Young Byun}}
\author[3,4]{\href{https://orcid.org/0000-0001-6558-9053}{Jongho Park}}
\author[3]{\href{https://orcid.org/0000-0002-0112-4836}{Sincheol~Kang}}
\author[1,3]{\href{https://orcid.org/0000-0002-9598-0018}{Naeun~Shin}}
\author[3]{\href{https://orcid.org/0000-0002-6269-594X}{Sang-Sung~Lee}}
\author[3]{\href{https://orcid.org/0000-0001-7003-8643}{Taehyun Jung}}

\affil[1]{Department of Physics and Astronomy, Seoul National University, Gwanak-gu, Seoul 08826, Republic of Korea; \email{mckam@snu.ac.kr, trippe@snu.ac.kr}}
\affil[2]{SNU Astronomy Research Center, Seoul National University, Gwanak-gu, Seoul 08826, Korea}
\affil[3]{Korea Astronomy and Space Science Institute, Daedeok-daero 776, Yuseong-gu, Daejeon 34055, Republic of Korea}
\affil[4]{Institute of Astronomy and Astrophysics, Academia Sinica, P.O. Box 23-141, Taipei 10617, Taiwan}

\def\corrauthor{
S. Trippe
}

\def\runningauthor{
Kam et al.
}

\def\runningtitle{
Crab Nebula as Polarization Angle Calibrator for KVN
}

\def\keywords{
polarization --- radiation mechanisms: non-thermal --- instrumentation: polarimeters
}

\def\abstracttext{
\noindent The Crab nebula is widely used as a polarization angle calibrator for single-dish radio observations because of its brightness, high degree of linear polarization, and well-known polarization angle over a wide frequency range. However, the Crab nebula cannot be directly used as a polarization angle calibrator for single-dish observations with the Korean VLBI Network (KVN), because the beam size of the telescopes is smaller than the size of the nebula. To determine the polarization angle of the Crab nebula as seen by KVN, we use 3C~286, a compact polarized extragalactic radio source whose polarization angle is well-known, as a reference target. We observed both the Crab nebula and 3C~286 with the KVN from 2017 to 2021 and find that the polarization angles at the total intensity peak of the Crab nebula (equatorial coordinates (J2000) R.A. $=$ 05$^{\rm h}$34$^{\rm m}$32.3804$^{\rm s}$ and Dec $=$ 22$^\circ$00'44.0982") are $154.2^\circ \pm 0.3^\circ$, $151.0^\circ \pm 0.2^\circ$, $150.0^\circ \pm 1.0^\circ$, and $151.3^\circ \pm 1.1^\circ$ at 22, 43, 86, and 94~GHz, respectively. We also find that the polarization angles at the pulsar position (RA $=$ 05$^{\rm h}$34$^{\rm m}$31.971$^{\rm s}$ and Dec $=$ 22$^\circ$00'52.06") are $154.4^\circ\pm 0.4^\circ$, $150.7^\circ\pm 0.4^\circ$, and $149.0^\circ\pm 1.0^\circ$ for the KVN at 22, 43, and 86~GHz. At 129 GHz, we suggest to use the values $149.0^\circ \pm 1.6^\circ$ at the total intensity peak and $150.2^\circ \pm 2.0^\circ$ at the pulsar position obtained with the Institute for Radio Astronomy in the Millimeter Range (IRAM) 30-meter Telescope. Based on our study, both positions within the Crab nebula can be used as polarization angle calibrators for the KVN single-dish observations.
}

\begin{document}
\jkashead 


\section{Introduction}\label{sec:intro}

Radio observatories with circular polarization receivers, such as the Korean VLBI Network (KVN), measure the electric vector position angle (EVPA) $\chi$ from the phase difference between the left circular polarization (LCP) and right circular polarization (RCP)  \citep{cotton1993}. However, an arbitrary offset in this phase difference is usually introduced by parts of the receiver such as synthesizer or sampler, which results in an offset in the EVPA. Since this offset tends to change when the system is turned on/off or when the frequency setting is modified, it is necessary to observe a source with known EVPA, an ``absolute EVPA calibrator'', to correct the phase difference and measure the intrinsic EVPA. 

A good absolute EVPA calibrator should be bright, spatially compact, and stable in polarization. Commonly, the AGN 3C~286 is considered as the source which best satisfies these three conditions \citep{Agudo+2012, Perley&Butler2013a, Perley&Butler2013b}. 3C~286 is a quasar with an angular size of a few arcseconds at centimeter wavelengths \citep{Akujor+1995}. The flux density $S_\nu$ from 1 to 50 GHz has been stable to within $\sim 1\%$ over 30 years \citep{Perley&Butler2013a} and exhibits a power-law spectrum with a spectral index $\alpha \sim -0.61$ (defined by $S_\nu \propto \nu^\alpha$) \citep{An+2017}. Likewise, the polarized flux is very stable, albeit showing a slow secular increase of less than 2\% per century at any given frequency \citep{Perley&Butler2013b}. 

The EVPA of 3C 286 has been studied across a wide frequency range. \citet{Bignell+1973} measured $\chi = 33.0^\circ \pm 0.9^\circ$ at 6.7~GHz and $\chi = 31.0^\circ \pm 1.3^\circ$ at 10.7~GHz with the 46-m Algonquin Radio Telescope. Based on this, \citet{Perley&Butler2013b} assumed that the EVPA of 3C~286 is 33$^\circ$ from 1 to 8~GHz and found that the EVPA rotates to $35.8^\circ \pm 0.1^\circ$ at 45~GHz. Polarization measurements with the IRAM Pico Veleta 30-meter Telescope (PV) \citep{Agudo+2012}, Submillimeter Array (SMA) \citep{Marrone2006, Hull+2016}, the Combined Array for Research in Millimeter-wave Astronomy (CARMA) \citep{Hull+2014, Hull&Plambeck2015}, and the Atacama Large Millimeter/submillimeter Array (ALMA) \citep{Nagai+2016} found that this trend continues to higher frequencies, with the EVPA increasing up to $41^\circ$. Likewise, the fractional polarization increases from 8.6\% to 17\% with increasing frequency \citep{Agudo+2012, Hull&Plambeck2015, Hull+2016, Nagai+2016}.  

Even though 3C 286 is compact and its EVPA is well-studied across a wide frequency range, it is not bright enough to be used as the absolute EVPA calibrator for the KVN which observes in the frequency range 22--129~GHz. The total flux density is 1.5~Jy at 49~GHz \citep{Perley&Butler2013a} and decreases to below 1~Jy at higher frequencies \citep{Agudo+2012, Hull&Plambeck2015, Hull+2016, Nagai+2016}. As a result, KVN cannot reliably detect the polarized flux at $\geq$ 86 GHz, and when it does, the resulting EVPA measurement comes with uncertainties larger than 5$^\circ$.

The Crab nebula, however, is much brighter than 3C 286. It is a supernova remnant powered by a relativistic wind from the pulsar located at its center \citep{Hester2008}. The integrated total flux of the Crab nebula is larger than 300~Jy throughout the KVN frequency range \citep{Macias-Perez+2010, Weiland+2011} and exhibits a power-law spectrum with a spectral index $\alpha \sim -0.3$ between 1 and 353~GHz \citep{Baars+1977, Macias-Perez+2010, Arendt+2011, Weiland+2011, Ritacco+2018} and $\alpha \sim -0.7$ between $10^4$ and $10^6$~GHz \citep{Macias-Perez+2010}. Excess emission observed above the synchrotron spectrum between $10^3$ and $10^4$~GHz is explained by thermal emission of dust at a temperature $T \sim 46$~K \citep{Strom+1992}. 

\citet{Aumont+2010} mapped the Crab nebula with the instruments XPOL installed at PV at 86 GHz. They measured an EVPA of $148.8^\circ \pm 0.2^\circ$ and a fractional polarization of $7.7\% \pm 0.2\%$ integrated over the entire nebula. \citet{Weiland+2011} observed the Crab nebula with WMAP at 22.70, 32.96, 40.64, 60.53, and 92.95~GHz for seven years and found an integrated EVPA of $149^\circ$--$150^\circ$ and a fractional polarization of $\sim 7\%$ across the observing frequency range. 

Even though the Crab nebula is one of the brightest radio sources and its integrated EVPA is well-studied, it cannot be directly used as an absolute EVPA calibrator for KVN because the angular size of the nebula is larger than the angular resolution of KVN. The average beam sizes of the KVN antennas are 129'', 63'', 31'', 29'', and 23'' at 22, 43, 86, 94, and 129~GHz, respectively,\footnote{\url{ https://radio.kasi.re.kr/status_report/files/KVN_Status_Report_2021.pdf}} while the angular size of the Crab nebula is 7$\times$5 arcminutes. High resolution polarization observations of the Crab nebula at millimeter wavelengths show that the EVPA varies considerably with position \citep{Aumont+2010, Ritacco+2018}. This implies that the EVPA integrated over the KVN beam will be different from the EVPA integrated over the entire nebula obtained from previous studies. In principle, one can infer the EVPA integrated over the KVN beam if there is a spatially resolved linear polarization image of the Crab nebula at each frequency. This is possible at 86 GHz because there is an XPOL map \citep{Aumont+2010}, but not at other frequencies. Therefore, it is essential to find the EVPA of the Crab nebula as measured with KVN to use it as an absolute EVPA calibrator for KVN.

One way to infer the EVPA of the Crab nebula measured with KVN is using 3C 286 that is compact and has stable, well-known EVPA across the KVN frequency range. The uncertainty in the EVPA of 3C 286 obtained from any single session is too large, but this uncertainty can be reduced by combining data from multiple observing sessions; this is permitted because the EVPA of 3C~286 is expected to be stable in time \citep{Perley&Butler2013b}.

We present the results of four years of polarization observations of the Crab nebula and 3C 286 obtained with KVN at 22, 43, 86, and 129~GHz. We also present ALMA data of 3C~286 from the ALMA Calibrator Source Catalogue\footnote{\url{https://almascience.eso.org/sc}} to check the stability of 3C 286 and use it to find the EVPA of the Crab nebula seen by KVN. KVN observations and ALMA archive data are summarized in Section~\ref{sec:obs}. The polarization of the Crab nebula and 3C 286 are presented in Section~\ref{sec:results}. In addition to the angular resolution, we discuss other factors that contribute to the variation of the EVPA with frequency in Section~\ref{sec:discussion}. A summary and conclusions of this study follow in Section~\ref{sec:summary}.

\begin{table}[t]
    \centering
    \setlength{\tabcolsep}{5pt}
    \caption{Log of Crab nebula observations, by station and frequency.}
    \begin{tabular}{lllll}
    \toprule
        Date & TN & US & YS & Crab \\
        \midrule 
        2017-02-24 & 86.2/--- & 22.4/43.0 & 86.2/--- & P \\ 
        2017-03-25 & --- & 22.4/43.0 & 86.2/129.4& P \\ 
        2017-04-23 & 86.2/129.4 & 22.4/43.0 & 86.2/129.4& P \\ 
        2017-05-31 & --- & 22.4/43.0 & 86.2/129.4 & P\\ 
        2017-09-24 & 86.2/129.4 & 22.4/43.0 & 86.2/129.4 & P\\ 
        2017-10-27 & 22.4/43.0 & 86.2/129.4 & 86.2/129.4 & P\\ 
        2017-11-19 & 22.4/43.0 & 22.4/43.0 & 86.2/129.4 & P\\ 
        2017-12-21 & 86.2/129.4 & 22.4/43.0 & 86.2/--- & P\\
        2018-02-14 & 22.4/43.0 & 22.4/43.0 & 86.2/129.4 & P\\ 
        2018-03-21 & 22.4/43.0 & --- & 86.2/--- & P\\ 
        2018-05-05 & --- & 22.4/43.0 & 86.2/--- & P\\ 
        2018-06-10 & 22.4/43.0 & --- & 86.2/--- & P\\
        2018-10-01 & 22.4/43.0 & 94.0/--- & 86.2/--- & P\\ 
        2018-11-02 & 22.4/43.0 & 94.0/141.0 & 86.2/129.4 & P\\ 
        2018-12-02 & --- & 94.0/--- & --- & P\\ 
        2019-01-04 & --- & --- & 86.2/129.4 & P\\ 
        2019-02-09 & 22.4/43.0 & 94.0/141.0 & 86.2/129.4 & P\\ 
        2019-03-03 & 22.4/43.0 & 22.4/43.0 & 86.2/129.4 & P\\ 
        2019-04-03 & 22.4/43.0 & 94.0/141.0 & 86.2/129.4 & P\\ 
        2019-04-08 & 22.4/43.0 & 94.0/141.0 & 22.4/43.0 & P\\
        2019-05-01 & 22.4/43.0 & 86.2/129.4 & 86.2/129.4 & P\\ 
        2019-06-01 & 22.4/43.0 & 94.0/141.0 & 86.2/129.4 & P\\ 
        2019-10-14 & 22.4/43.0 & --- & 86.2/129.4 & P\\ 
        2019-11-26 & 22.4/43.0 & --- & 86.2/129.4 & P, I\\ 
        2019-12-15 & 22.4/43.0 & --- & 86.2/129.4 & P, I\\ 
        2019-12-19 & 22.4/43.0 & 86.2/129.4 & 86.2/129.4 & P, I \\ 
        2020-01-19 & 22.4/43.0 & --- & 86.2/129.4 & P, I   \\
        2020-02-22 & 22.4/43.0 & 86.2/129.4 & 86.2/129.4 & P, I   \\
        2020-03-20 & 22.4/43.0 & 86.2/129.4 & 86.2/129.4 & P, I   \\
        2020-04-18 & 22.4/43.0 & 86.2/129.4 & 86.2/129.4 & P, I   \\
        2020-05-24 & --- & 22.4/43.0 & ---  & P, I \\
        2020-12-08 & 22.4/43.0 & 86.2/129.4 & ---  & P, I \\
        2021-01-03 & 22.4/43.0 & 86.2/--- & 22.4/43.0 & I \\ 
        2021-01-22 & 22.4/43.0 & 22.4/43.0 & 86.2/129.4 & I \\ 
        2021-02-10 & 22.4/43.0 & --- & 86.2/129.4 & I \\ 
        2021-02-13 & 22.4/43.0 & 94.0/141.0 & 86.2/129.4 & I \\ 
    \bottomrule
    \end{tabular}
    \tabnote{
        Frequencies are in units of GHz. `P' and `I' represent the pulsar position and total intensity peak position, respectively. 
    }
    \label{tab1}
\end{table}

\section{Observations and Data Reduction}\label{sec:obs}

\subsection{KVN}\label{sec:kvn}

Both the Crab nebula and 3C 286 have been used as absolute EVPA calibrators for the Plasma-physics of Active Galactic Nuclei (PAGaN) blazar monitoring project, a KVN Key Science Project, started in 2017 \citep{Park+2018}. Single-dish observations of these sources are conducted close to (mostly within 1$\sim$2 days) each VLBI observing run for the EVPA calibration. Each station operates at two frequencies simultaneously in dual-polarization mode \citep{Lee+2011}. KVN Yonsei (YS) and KVN Tamna (TN) have mostly been used at 86/129~GHz and 22/43~GHz, respectively. KVN Ulsan (US) has observed at either 22/43 GHz or 86/129 GHz depending on the weather condition. From November 2018 to June 2019, US was mostly used at 94/141 GHz. In this study, we included PAGaN runs from February 2017 to February 2021; the observing dates and frequencies are summarized in Table~\ref{tab1}. 

Observations are executed in scans of 40 minutes. Each scan consists of cross-scan pointing, sky dipping curve measurement, and 8 sets of position-switching cycles described in \citet{Kang+2015}. Each position-switching cycle alternately observes the target source and a reference sky patch. We adopt the standard deviation of the EVPA values obtained from the 8 sets of position switching measurements as the random error of the EVPA. Jupiter and the radio galaxy 3C~84 are assumed to be unpolarized on arcsecond scales, and we observed at least one of them in each session for the purpose of instrumental polarization calibration. We reduced the data with the KVN single-dish data processing pipeline described in \citet{Kang+2015}. We allocated at least two scans to each of the Crab nebula and 3C 286 in each session to increase the signal-to-noise ratio. 

For the Crab nebula, we observed either the total intensity peak at 150 GHz located at equatorial coordinates (J2000) R.A. $=$ 05$^{\rm h}$34$^{\rm m}$32.3804$^{\rm s}$ and Dec $=$ 22$^\circ$00'44.0982" \citep{Ritacco+2018} or the pulsar position at RA $=$ 05$^{\rm h}$34$^{\rm m}$31.971$^{\rm s}$ and Dec $=$ 22$^\circ$00'52.06". From November 2019 to December 2020, we observed each time both positions to investigate the EVPA difference between them. Either position was observed two to four times at elevations between $30^\circ$ and $70^\circ$ where the normalized gain becomes larger than 0.8. \footnote{\url{https://radio.kasi.re.kr/status_report/files/KVN_status_report_2021.pdf}}

\subsection{ALMA}

The ALMA Calibrator Source Catalogue provides linear polarization measurements of 3C~286 obtained with the 7-m Array (Misato Fukagawa, priv. commun.). To confirm the stability of 3C 286 and to use them as reference values for finding the EVPA of the Crab nebula seen by KVN, we collected the 3C~286 data obtained from January 18, 2019 (MJD 58501) to December 8, 2021 (MJD 59556); see Figure~\ref{fig1}. The data were obtained in Band 3 (91.5 and 103.5 GHz), Band 6 (233~GHz), and Band 7 (343.5 and 349.5~GHz). In case of Band 7, we combined the data at 343.5 and 349.5 GHz and used the geometrical mean of the two frequencies, 346.5 GHz, as proxy for the observing frequency.

\section{Results}\label{sec:results}

\subsection{The Stability of 3C 286}

Before using the EVPA of 3C 286 to find $\chi_{\rm obs}$ for the total intensity peak of the Crab nebula (hereafter I-peak), we first need to check its stability. The EVPA of 3C 286 has been assumed to be stable based on the fact that the total and polarized flux from 1 to 50~GHz have been stable for over 30 years \citep{Perley&Butler2013a, Perley&Butler2013b}. This allows us to compare the EVPA of 3C 286 measured with KVN with that measured with VLA 10--20 years ago \citep{Perley&Butler2013b} at $\leq 50$ GHz. At higher frequencies, \citet{Agudo+2012} measured the EVPA of 3C~286 with PV from September 2006 to January 2012 and showed that the EVPA of 3C~286 is stable for $\sim$5 years at 86 GHz and $\sim$2 years at 230 GHz. Since this is shorter than the time gap of 10--15 years between the KVN and PV observing periods, we used ALMA data to test the stability of 3C 286 at $\geq 86$ GHz over a longer time scale. 

\begin{figure}[t]
\centering 
\includegraphics[width=\columnwidth]{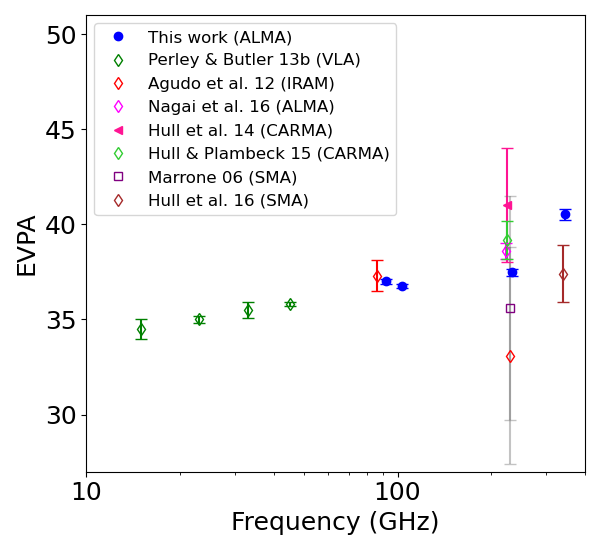}
\caption{The EVPA of 3C~286 measured with ALMA (this work), VLA \citep{Perley&Butler2013b}, IRAM \citep{Agudo+2012}, SMA \citep{Marrone2006, Hull+2016}, CARMA \citep{Hull+2014, Hull&Plambeck2015}, and ALMA \citep{Nagai+2016} are shown for reference.}
\label{fig:crab1_pa}\label{fig1}
\end{figure}

The EVPA and fractional polarization of 3C~286 measured with ALMA do not show any systematic trend with time (see Appendix~\ref{sec:3c286}). In addition, we found that the standard deviations of the EVPA are close to the mean of the measurement uncertainties at all four ALMA frequencies. This suggests that the EVPA of 3C 286 measured with ALMA is stable over the 3-year observing period at frequencies up to 346.5 GHz. In addition, the weighted mean and its standard error, $37.0^\circ \pm 0.1^\circ$ at 91.5~GHz, agrees with the value $37.3^\circ \pm 0.8^\circ$ measured with PV at 86 GHz \citep{Agudo+2012}, despite the time gap of $\sim10$ years between the two datasets (Figure~\ref{fig1}). This suggests that the EVPA of 3C~286 has been stable for more than a decade at $\sim$86~GHz.
This enables us to estimate the EVPA of the Crab nebula seen by KVN by comparing the EVPA of 3C~286 measured with KVN (see Appendix~\ref{sec:3c286}) at 86 GHz with that measured with PV at 86 GHz 10--15 years ago. \

\begin{table}[t]
    \centering
    \caption{Our EVPA values for the Crab nebula (both I-peak and pulsar position) for KVN, along the reference values for 3C~286.}
    \begin{tabular}{c c c c}
    \toprule 
    $\nu$ [GHz] & $\chi_{\rm ref,3C286}$ [$^\circ$] & $\chi_{\rm peak}$ [$^\circ$] &  $\chi_{\rm pulsar}$ [$^\circ$] \\
    \midrule 
    22.4 & 35.0 $\pm$ 0.2\tnote{a} & $154.2 \pm 0.3$ & 154.4 $\pm$ 0.4    \\
    43.0 & 35.8 $\pm$ 0.1\tnote{b} & $151.0 \pm 0.2$ & 150.7 $\pm$ 0.4  \\
    86.2 & 37.3 $\pm$ 0.8\tnote{c} & $150.0 \pm 1.0$ & 149.0 $\pm$ 1.0  \\ 
    94.0 & 37.0 $\pm$ 0.1\tnote{d} & $151.3 \pm 1.1$ & ---  \\ 
    129.4 & --- & $149.0 \pm 1.6$\tnote{e} & 150.2 $\pm$ 2.0\tnote{e}  \\ 
    \bottomrule
    \end{tabular}
    \tabnote{ 
        $^{\rm a}$ VLA 23 GHz \citep{Perley&Butler2013b} \\ 
        $^{\rm b}$ VLA 45 GHz \citep{Perley&Butler2013b} \\ 
        $^{\rm c}$ XPOL 86 GHz \citep{Agudo+2012} \\ 
        $^{\rm d}$ ALMA 91.5 GHz (this work) \\ 
        $^{\rm e}$ XPOL 86 GHz \citep{Ritacco+2018}
    }
    \label{tab2}
\end{table}\

\subsection{The EVPA of the Crab Nebula}

Given that the EVPA measured with circular polarization receivers has an unknown systematic offset $\chi_{\rm off}$ introduced by parts of the receiver system, what we measure is the EVPA difference $\Delta\chi$ between 3C 286 and the Crab nebula:
\begin{equation}\label{eq:1}
    \begin{aligned}
    \Delta\chi & = (\chi_{\rm off} + \chi_{\rm 3c286}) - (\chi_{\rm off} + \chi_{\rm crab}) \\ 
    & = \chi_{\rm 3c286} - \chi_{\rm crab}
    \end{aligned}
\end{equation}

\noindent where $\chi_{\rm 3c286}$ and $\chi_{\rm crab}$ are the EVPA values of 3C 286 and the Crab nebula respectively. The measurement uncertainties in $\chi_{\rm crab}$ are negligible compared to those of 3C 286 because the Crab nebula is much brighter than 3C 286. Therefore, we assume those to be negligible so that we can convert $\Delta\chi$ to $\chi_{\rm 3c286}$. Then, we can determine $\chi_{\rm crab}$ that makes $\langle\Delta\chi\rangle - \chi_{\rm crab}$ equal to the known EVPA values of 3C 286 at each frequency, where $\langle\Delta\chi\rangle$ is the weighted mean of $\Delta\chi$ observed in multiple epochs at each frequency. 
We adopt the standard errors of mean of the EVPA of 3C 286 as the uncertainties of $\chi_{\rm crab}$ because we neglected the EVPA uncertainties of the Crab nebula as they are much smaller than those of 3C 286.

\begin{figure}[t]
\centering 
\includegraphics[width=\columnwidth]{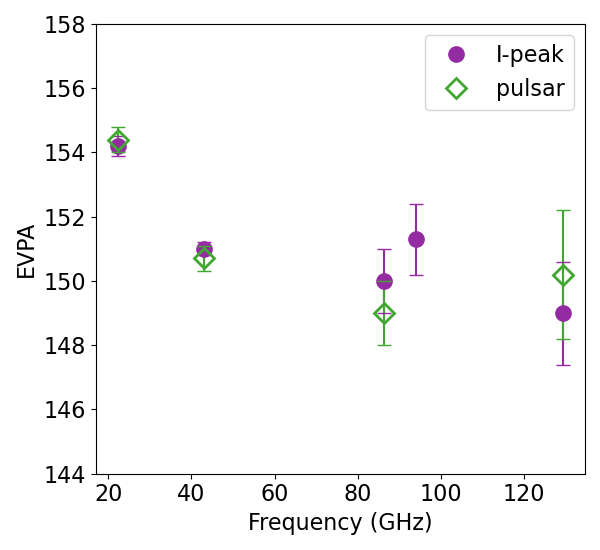}
\caption{The EVPA of I-peak and pulsar position of the Crab nebula as observed by KVN.}
\label{fig:crab1_pa}\label{fig2}
\end{figure}

The known EVPA values of 3C 286 at the frequencies closest to the KVN frequencies are $35.0^\circ \pm 0.2^\circ$ and $35.8^\circ \pm 0.1^\circ$ at 23 and 45 GHz \citep{Perley&Butler2013b}, $37.3^\circ \pm 0.8^\circ$ at 86 GHz \citep{Agudo+2012}, and $37.0^\circ \pm 0.1^\circ$ and $36.8^\circ \pm 0.3^\circ$ at 91.5 and 103.5 GHz (from the ALMA data), respectively. Using these values, $\chi_{\rm obs}$ that we obtained for the Crab nebula I-peak are 154.2$^\circ$ $\pm$ 0.3$^\circ$, 151.0$^\circ$ $\pm$ 0.2$^\circ$, 150.0$^\circ$ $\pm$ 1.0$^\circ$, and 151.3$^\circ$ $\pm$ 1.1$^\circ$ at 22, 43, 86, and 94 GHz (Table~\ref{tab2} and Figure~\ref{fig2}). Additionally, we measured the EVPA at the pulsar position relative to the EVPA at I-peak (Table~\ref{tab3}), eventually resulting in EVPA values for the pulsar position of 154.4$^\circ$ $\pm$ 0.4$^\circ$, 150.7$^\circ$ $\pm$ 0.4$^\circ$, and 149.0$^\circ$ $\pm$ 1.0$^\circ$ at 22, 43, and 86 GHz, respectively (Table~\ref{tab2} and Figure~\ref{fig2}). The EVPA difference between the two positions is less than $1^\circ$ across the KVN frequency range (Table~\ref{tab3}). This observation is in fair agreement with previous studies \citep{Ritacco+2018}. 

At 129~GHz, the number of EVPA measurements of 3C 286 with KVN is too small to determine $\chi_{\rm obs}$ for the I-peak reliably. Considering that the angular resolution of XPOL at 86~GHz, 27'', is close to that of KVN at 129~GHz, 23'', we adopted the values $149.0^\circ \pm 1.6^\circ$ and $150.2^\circ \pm 2.0^\circ$ obtained with XPOL at 86 GHz \citep{Ritacco+2018} as the reference values for the Crab I-peak and pulsar position at 129 GHz, respectively. 

\begin{table}[t]
    \centering
    \caption{The mean and standard error of the EVPA at I-peak with respect to the EVPA at the pulsar position.}
    \begin{tabular}{c c c c c}
    \toprule
    $[^\circ]$ & 22 GHz & 43 GHz & 86 GHz & 129 GHz \\
    \midrule
    $\chi_{\rm rel}$ & -0.27$\pm$0.23 & 0.28$\pm$0.34 & 0.98$\pm$0.31 & 0.52$\pm$0.59 \\
    \bottomrule
    \end{tabular}
    \label{tab3}
\end{table}

\subsection{The Fractional Polarization of the Crab Nebula}

The fractional polarization of the Crab nebula tends to increase with frequency at both positions (Table~\ref{tab4}). This is consistent with previous observations which found that the fractional polarization is larger at the central region where the pulsar and the I-peak are located \citep{Aumont+2010, Ritacco+2018}. The fractional polarization measured with KVN at 86 and 129~GHz is 1\%--2\% lower than that measured with the PV instruments XPOL at 86~GHz and NIKA at 150~GHz even though their angular resolutions and frequencies are similar (see Appendix~\ref{sec:polar}). This is currently under investigation and will be discussed in a future study.
Previous studies using POLKA \citep{Wiesemeyer+2014}, XPOL \citep{Aumont+2010}, SCUPOL \citep{Matthews+2009}, and NIKA \citep{Ritacco+2018} found that the EVPA at I-peak and at the pulsar position are almost identical within the measurement uncertainties, even though the absolute values are slightly different since the instruments operated at different frequencies \citep[see Table 1 in][]{Ritacco+2018}. \

\begin{table}[t]
    \centering
    \setlength{\tabcolsep}{10pt}
    \caption{Fractional polarization at the pulsar position and I-peak of the Crab nebula. }
    \begin{tabular}{c c c}
    \toprule
    $\nu$ [GHz] & Pulsar [\%] & I-peak [\%]  \\
    \midrule
    22.4 & 12.62 $\pm$ 0.01 & 12.86 $\pm$ 0.01  \\
    43.0 & 14.18 $\pm$ 0.01 & 13.39 $\pm$ 0.01 \\
    86.2 & 15.64 $\pm$ 0.04 & 17.89 $\pm$ 0.04 \\
    129.4 & 15.75 $\pm$ 0.20 & 16.36 $\pm$ 0.14 \\
    \bottomrule
    \end{tabular}
    \label{tab4}
\end{table}

\section{Discussion}\label{sec:discussion}

Our analysis has reliably measured the EVPA for the Crab nebula, both at I-peak and the pulsar position, across the frequency range accessible to KVN observations. The values we provide here can serve as a calibration reference for current and future polarization studies based on KVN data.

The EVPA values at all frequencies and positions are located near 150$^\circ$. Notably, the EVPA values of the Crab nebula (at both positions) at 129 GHz, which we adopted from XPOL results, are consistent with the KVN values at 43, 86, and 94 GHz within twice the measurement uncertainty. Possible exceptions are the EVPA values at 22 GHz; $154.2^\circ\pm0.3^\circ$ at the I-peak position and $154.4^\circ\pm0.4^\circ$ at the pulsar position deviate from the EVPA values at the rest of the frequencies by ten times the measurement uncertainty. Here we explore mechanisms that might introduce a frequency dependency and the resulting limits on any feasible systematic trend.  \

\subsection{Angular Resolution}

The EVPA of the Crab nebula is distributed around $140^\circ$--$150^\circ$ in the vicinity of I-peak, and changes by $30^\circ$ to $100^\circ$ toward the edge of the nebula \citep[see Figure~3 in][]{Ritacco+2018}. Therefore, different EVPA values will be observed with different spatial resolutions even when observing the same position. Indeed, the EVPA at I-peak obtained with KVN at 86 GHz with 32'' angular resolution, $150.0^\circ \pm 1.0^\circ$, is consistent with the value $149.0^\circ \pm 1.6^\circ$ obtained with XPOL at 86 GHz with 27'' angular resolution. However, the difference between values measured with KVN and XPOL increases with increasing KVN beam size. This suggests that the spatial resolution may noticeably affect the observed EVPA value as function of frequency.  

One more thing we should take into account is the expansion of the Crab nebula. It is expanding at 1500~km\,s$^{-1}$ which corresponds to 0.15'' per year at a distance of 2~kpc \citep{Bietenholz+1991b}. This corresponds to an angular expansion of about 1.5'' since the XPOL map was obtained in January 2009. This would be negligible for the measurements for the entire nebula which extended over 7 by 5 arcminutes. The beam of KVN is much smaller than the angular size of the Crab nebula. The smallest beam size of KVN is 23'' at 129 GHz. The expansion of 1.5'' corresponds to $\sim$0.4\% of the KVN 129~GHz beam in area. Therefore, we conclude that the effect of the expansion is negligible when we compare the EVPA values obtained with KVN and IRAM. \

\subsection{Spectral Index}

\citet{Arendt+2011} mapped the Crab nebula with the GISMO bolometer camera on PV and constructed spectral index maps at a resolution of 20'' using the wavelength pairs (i) 2~mm and 6~cm, and (ii) 2~mm and 21~cm. For both frequency pairs, the spectral index becomes closer to zero toward the center of the nebula, with significant substructure on angular scales smaller than the KVN beam sizes. This implies that the integrated EVPA within the beam may change, depending on the observing frequency. \citet{Ritacco+2018} collected maps of the Crab nebula obtained from different instruments at different frequencies, and degraded them to the 27'' resolution of XPOL to remove the resolution effect from the observed EVPA. The EVPA at the I-peak and the pulsar position differs depending on the frequency even though the resolutions are the same \citep[see Table~1 in][]{Ritacco+2018}. This suggests that spectral index variations may contribute to variations of the EVPA with frequency, if any.  \

\subsection{Instrumental Polarization}

An ideal circular polarization receiver responds to only one desired polarization. However, a real circular polarization receiver also responds to the undesired orthogonal polarization, resulting in instrumental polarization \citep{Roberts1994}. The instrumental polarization consists of on-axis and off-axis components. The on-axis instrumental polarization is a component at the center of the antenna beam that is primarily caused by the polarization splitter which splits the received signal into RCP and LCP \citep{Napier1999}. This can be treated as constant across the beam and can be removed by standard instrumental polarization calibration processes. On the other hand, the off-axis instrumental polarization is caused by (i) the reflection at the surface of the paraboloid main reflector \citep{Rudge+1982, Napier1999} and (ii) complicated quasi-optics for simultaneous multi-frequency observation, and their contribution varies across the beam.

The response function to the desired polarization, the co-polarization beam, has a circular symmetry with a maximum at the center. On the other hand, the response function to the off-axis instrumental polarization, the cross-polarization beam, has a vertical symmetry with two primary beams, mainly due to the contribution of ellipsoidal mirrors in the quasi-optics. 
The amplitude of the cross-polarization beam is smaller than that of the co-polarization at the center by more than 50 dB, indicating that the off-axis instrumental polarization does not affect observations of compact objects such as blazars. However, it might need to be considered for observations of sources which are larger than the primary cross-polarization beams at all frequencies of KVN, such as the Crab nebula, because the two primary beams of cross-polarization reach their peak values at approximately the location of the half power of the primary co-polarization beam. 

One way to test the effect of cross-polarization on $\chi_{\rm obs}$ for the Crab nebula is convolving the XPOL map\footnote{\url{ https://wiki.cosmos.esa.int/planck-legacy-archive/index.php/External_maps}} \citep{Aumont+2010}, the only available map of the Crab nebula at KVN frequencies, with the co-polarization and cross-polarization beams of KVN, and comparing the total flux values obtained with the two beams at I-peak. We confirmed that the ratio between the flux obtained with cross-polarization beam and that obtained with co-polarization beam is less than 0.015\% at 22--86 GHz. According to the error propagation of polarization, this flux ratio corresponds to the EVPA of 0.03$^\circ$ at all KVN freuqencies. Therefore, we conclude that the effect of the instrumental polarization is negligible. \

\subsection{Faraday Rotation}

When linearly polarized radiation passes through a magnetized plasma, the EVPA rotates due to Faraday rotation. The amount of Faraday rotation is proportional to the integral of the product of the electron density $n_{\rm e}$ and magnetic field $\vec{B}$ along the line of sight $\vec{l}$, the rotation measure (RM)
\begin{equation}
    {\rm RM} \propto \int n_{\rm e}\vec{B}\cdot d\vec{l} 
\end{equation}
\begin{equation}
    \chi_{\rm obs} = \chi_{0} + \lambda^2 \,\rm RM 
\end{equation}
where $\chi_{\rm obs}$ and $\chi_{0}$ are the observed and intrinsic EVPA, respectively. The large-scale RM over the Crab nebula is between $-21$~rad\,m$^{-2}$ and $-43$~rad\,m$^{-2}$ from centimeter to millimeter wavelengths \citep{Burn1966, Manchester1971, Rankin+1988, Bietenholz+1991a, Weiland+2011}. This corresponds to Faraday rotation of less than $0.5^\circ$ between 22 and 129 GHz, indicating that the effect of Faraday rotation is negligible within the KVN frequency range. \

\section{Summary and Conclusions}\label{sec:summary}

In this paper, we have reported the first cross-calibrated EVPA measurements of the Crab nebula at its total intensity peak (I-peak) and pulsar positions by KVN.

In order to obtain the intrinsic EVPA of the Crab nebula as observed by KVN, we used the EVPA of the standard calibrator 3C~286 as reference.

We first tested its stability using archival ALMA polarization measurements. The EVPA and fractional polarization of 3C~286 measured with ALMA between January 2019 and December 2021 are stable up to frequencies of 346.5 GHz at least within the 3-year observation period. The average EVPA at each ALMA frequency is consistent with the values obtained with other instruments over a wide frequency range. In particular, the value $37.0^\circ \pm 0.1^\circ$ measured with ALMA at 91.5~GHz is virtually identical to $37.3^\circ \pm 0.8^\circ$ obtained with the Pico Veleta observatory at 86~GHz between September 2006 and January 2012. This suggests that the EVPA of 3C~286 is stable for more than a decade at least for the KVN frequency range, and enable us to use it to estimate the EVPA of the Crab nebula seen by KVN. 

Using the EVPA values of 3C 286 as reference, we have shown that the EVPA at the Crab nebula I-peak seen by KVN is $154.2^\circ \pm 0.3^\circ$, $151.0^\circ \pm 0.2^\circ$, $150.0^\circ \pm 1.0^\circ$, and $151.3^\circ \pm 1.1^\circ$ at 22, 43, 86, and 94 GHz, respectively.
Using our measurements of the EVPA difference between the I-peak and the pulsar position, we have also found that the EVPA values at the pulsar position are $154.4^\circ \pm 0.4^\circ$, $150.7^\circ \pm 0.4^\circ$, and $149.0^\circ \pm 1.0^\circ$ at 22, 43, and 86 GHz respectively. Due to the lack of EVPA measurements of 3C~286 at 129 GHz, we adopted the XPOL values $149.0^\circ \pm 1.6^\circ$ and $150.2^\circ \pm 2.0^\circ$ for the Crab I-peak and pulsar position at 129 GHz, respectively, considering the fact that the resolution and frequency of XPOL at 86~GHz are close to that of KVN at 129~GHz. Indeed, the values obtained with XPOL are in agreement with that obtained with KVN at 43, 86, and 94 GHz, except for the values at 22 GHz. 

There are several factors that might contribute to a variation of the EVPA with frequency. Since the angular size of the Crab nebula is larger than the beams of the KVN, the integrated EVPA within the beam of KVN could be different at different frequencies. In addition, the spectral index is steeper in the outer region of the Crab nebula, which also may affect the observed EVPA at different frequencies. When considering the effects of instrumental polarization and Faraday rotation, we found them to be negligible.

Our analysis has demonstrated that the Crab nebula, which is a well-known EVPA calibrator in radio astronomy, can be used for this purpose also by the KVN. The specific (frequency-dependent) EVPA values we find can serve as references for current and future polarimetric studies with the KVN.

\begin{figure*}[t]
\centering 
\includegraphics[width=\textwidth]{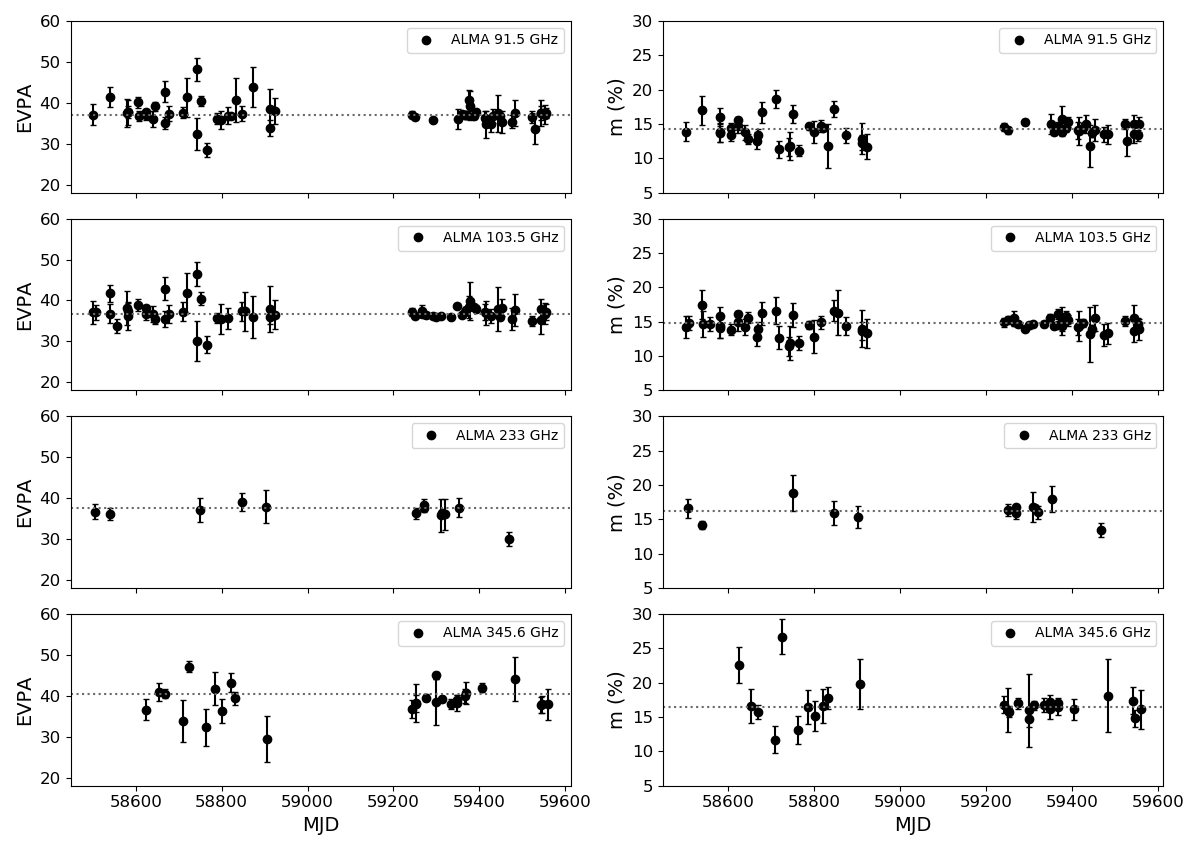}
\caption{The EVPA and fractional polarization of 3C 286 measured with ALMA. Gray-dotted lines on the left represent the weighted mean of the EVPA at each frequency. Gray-dotted lines on the right represent the weighted mean of the fractional polarization at each frequency.}
\label{fig:3c286_alma}\label{fig3}
\end{figure*}


\acknowledgments
We are grateful to the staff of the KVN who helped to operate the array and to correlate the data. We thank Jonathan Aumont and Alessia Ritacco for providing the XPOL data and useful discussions. PAGaN is a KVN Key Science Program. The KVN is a facility operated by the Korea Astronomy and Space Science Institute (KASI). The KVN observations and correlations are supported through the high-speed network connections among the KVN sites provided by the Korea Research Environment Open NETwork (KREONET), which is managed and operated by the Korea Institute of Science and Technology Information (KISTI). We acknowledge financial support from the National Research Foundation of Korea (NRF) grant 2022R1F1A1075115. J.P. acknowledges ﬁnancial support from the Korean National Research Foundation (NRF) via Global PhD Fellowship grant 2014H1A2A1018695 and support through the EACOA Fellowship awarded by the East Asia Core Observatories Association, which consists of the Academia Sinica Institute of Astronomy and Astrophysics, the National Astronomical Observatory of Japan, Center for Astronomical Mega-Science, Chinese Academy of Sciences, and the Korea Astronomy and Space Science Institute. This paper makes use of the following ALMA data: ADS/JAO.ALMA\#2011.0.00001.CAL. ALMA is a partnership of ESO (representing its member states), NSF (USA) and NINS (Japan), together with NRC (Canada), MOST and ASIAA (Taiwan), and KASI (Republic of Korea), in cooperation with the Republic of Chile. The Joint ALMA Observatory is operated by ESO, AUI/NRAO and NAOJ.


\appendix

\section{The Stability of 3C 286}\label{sec:3c286}

To determine if there is any systematic trend with time in the polarization of 3C 286 over the 3-year observing period of ALMA, we compared the standard deviations of the EVPA and the fractional polarization with the average values of their measurement uncertainties at each frequency. We found that the standard deviations of the EVPA and the fractional polarization are close to the mean of the measurement uncertainties at all four ALMA frequencies, suggesting that both the EVPA and the fractional polarization are stable at least for 3 years up to 346.5 GHz (see Figure~\ref{fig3} and Table~\ref{tab5}).

\begin{table}[t]
    \centering
    \setlength{\tabcolsep}{4pt}
    \caption{Polarization measurements of 3C 286 with ALMA. }
    \begin{tabular}{c c c c c c c}
    \toprule
     $\nu$ & $\chi$ & m & 
     s$_{\chi}$\tnote{a} & $\langle \varepsilon_{\chi} \rangle$\tnote{b} &
     s$_{\rm m}$\tnote{c} & $\langle \varepsilon_{\rm m} \rangle$\tnote{d} \\ 
    $[$GHz$]$ & [$^\circ$] & [\%] & [$^\circ$] & [$^\circ$] & [\%] & $[\%]$ \\
    \midrule
    91.5 & $37.0 \pm 0.1$ & $14.3 \pm 0.1 $ & 2.8 & 2.1 & 1.5 & 1.2 \\
    103.5 & $36.8 \pm 0.1$ & $14.7 \pm 0.1$ & 2.4 & 2.2 & 1.2 & 1.3 \\
    233 & $37.5 \pm 0.2$ & $16.3 \pm 0.2$ & 2.1 & 2.2 & 1.3 & 1.3 \\
    346.5 & $40.5 \pm 0.3$ & $16.5 \pm 0.2$ & 3.6 & 2.6 & 2.7 & 2.0 \\ 
    \bottomrule
    \end{tabular}
    \tabnote{ 
        $^{\rm a}$ Standard deviation of the EVPA. \\ 
        $^{\rm b}$ Mean EVPA error. \\
        $^{\rm c}$ Standard deviation of the fractional polarization. \\ 
        $^{\rm d}$ Mean fractional polarization error. \\ 
        } 
    \label{tab5}
\end{table}\

\begin{table}[t]
    \centering
    \caption{Fractional polarization measurements of 3C 286.}
    \begin{tabular}{c c c}
    \toprule 
     $\nu [$GHz$]$ & m$_{\rm kvn}$ [\%] & m$_{\rm ref}$ [\%] \\
    \midrule 
    22.4 & 10.5 $\pm$ 0.1 & 12.6\tnote{a}   \\
    43.0 & 10.9 $\pm$ 0.1 & 13.2\tnote{b}   \\
    86.2 & 11.3 $\pm$ 0.2 & 13.5 $\pm$ 0.3\tnote{c}  \\
    94.0 & 12.6 $\pm$ 0.4 & 14.3 $\pm$ 0.1\tnote{d}   \\  
    129.3 & 10.9 $\pm$ 0.8 & 14.7 $\pm$ 0.1\tnote{e}   \\ 
    \bottomrule 
    \end{tabular}
    \tabnote{ 
        $^{\rm a}$ VLA 22.4 GHz \citep{Perley&Butler2013b}  \\ 
        $^{\rm b}$ VLA 43.5 GHz \citep{Perley&Butler2013b}  \\ 
        $^{\rm c}$ XPOL 86 GHz \citep{Agudo+2012} \\ 
        $^{\rm d}$ ALMA 91.5 GHz (this work) \\
        $^{\rm e}$ ALMA 103.5 GHz (this work)
     }
    \label{tab6}
\end{table}\

\begin{figure*}[t]
\centering 
\includegraphics[scale=0.6]{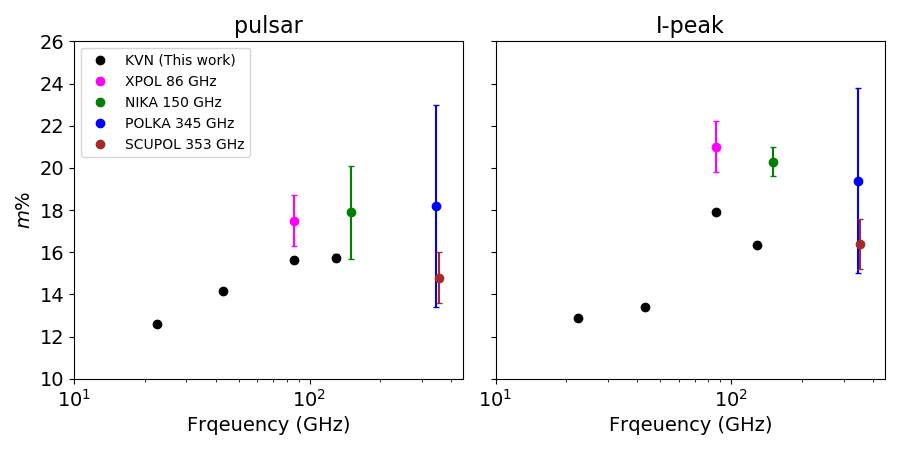}
\caption{Fractional polarization at pulsar and I-peak. XPOL, NIKA, POLKA, SCUPOL results are taken from Table~1 in \citet{Ritacco+2018}.
}  
\label{fig4}
\end{figure*}

\begin{figure}[t]
\centering 
\includegraphics[width=\columnwidth]{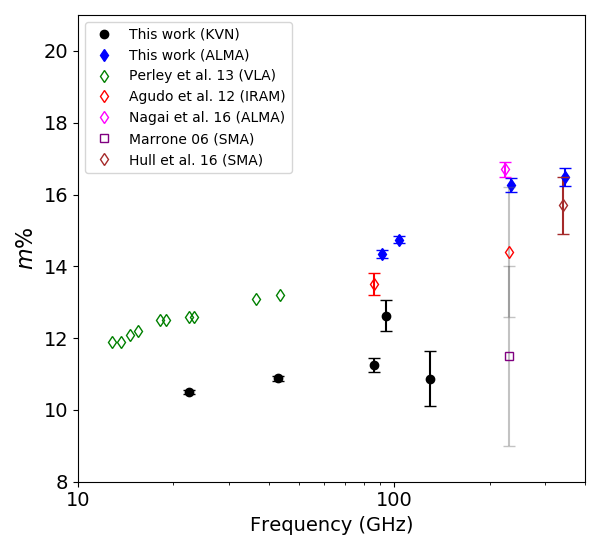}
\caption{The fractional polarization of 3C 286 obtained with KVN (this work), ALMA (this work), VLA \citep{Perley&Butler2013b}, IRAM \citep{Agudo+2012}, SMA \citep{Marrone2006, Hull+2016}, and ALMA \citep{Nagai+2016}.}
\label{fig5}
\end{figure}

\section{Fractional Polarization}\label{sec:polar}

The fractional polarization at the I-peak and pulsar positions of the Crab nebula are presented in Figure~\ref{fig4}. The beam sizes of KVN at 86 and 129~GHz are 32'' and 23'', respectively. These are very close to the 27'' and 18'' beam sizes of XPOL at 86~GHz and NIKA at 150~GHz, respectively. The apertures of the KVN and IRAM Pico Veleta antennas are 21~m and 30~m, respectively. The KVN antennas are shaped Cassegrain reflectors whose aperture efficiencies, compared to the usual combination of a parabolic main and a hyperbolic sub-reflector, are increased by slightly modifying the reflector surfaces \citep{Kim+2011}; the IRAM telescope uses a Nasmyth-focus configuration \citep{Baars+1987}. This makes the angular resolution of KVN close to that of IRAM at a given frequency, despite the different apertures.

Nevertheless, the fractional polarization of 16\%--18\% measured with KVN at 86.2 and 129.4~GHz (see Figure~\ref{fig4}) is somewhat lower than the values 17\%--21\% measured with XPOL at 86~GHz and NIKA at 150~GHz. This discrepancy in the fractional polarization corresponds to a loss of polarized flux of 11\%--14\% at 86~GHz and 12\%--20\% at 129~GHz. Likewise, the fractional polarization of 3C~286 measured with KVN is lower than that measured with other instruments (see Figure~\ref{fig5} and Table~\ref{tab6}). In terms of polarized emission, the loss corresponds to 12\%--17\% at 22--94 GHz and $26\%$ at 129~GHz. \\


\end{document}